\documentclass[conference]{IEEEtran}
\IEEEoverridecommandlockouts
\usepackage{cite}
\usepackage{amsmath,amssymb,amsfonts}
\usepackage{algorithmic}
\usepackage{graphicx}
\usepackage{textcomp}
\usepackage{xcolor}
\def\BibTeX{{\rm B\kern-.05em{\sc i\kern-.025em b}\kern-.08em
    T\kern-.1667em\lower.7ex\hbox{E}\kern-.125emX}}

\usepackage{eso-pic}
\usepackage{url}
\AddToShipoutPictureBG*{
  \AtPageUpperLeft{%
    \put(0,-40){\raisebox{15pt}{\makebox[\paperwidth]{\begin{minipage}{21cm}\centering
      \textcolor{gray}{This article has been accepted for publication in the proceedings of the \\2023 IEEE International Workshop on Advances in Sensors and Interfaces (IWASI).} 
    \end{minipage}}}}%
  }
  \AtPageLowerLeft{%
    \raisebox{25pt}{\makebox[\paperwidth]{\begin{minipage}{21cm}\centering
      \textcolor{gray}{ \copyright 2023 IEEE.  Personal use of this material is permitted.  Permission from IEEE must be obtained for all other uses, in any current or future media, including reprinting/republishing this material for advertising or promotional purposes, creating new collective works, for resale or redistribution to servers or lists, or reuse of any copyrighted component of this work in other works.
      }
    \end{minipage}}}%
  }
}
\begin{document}

\title{A Fast and Accurate Optical Flow Camera for Resource-Constrained Edge Applications\\
\thanks{This work was supported by the European Union’s ERA-NET CHIST-ERA 2018 Research and Innovation Programme APROVIS3D under the Grant SNF-19 20CH21\_186991.}
}

\author{\IEEEauthorblockN{Jonas Kühne, Michele Magno, Luca Benini}
\IEEEauthorblockA{\textit{Dept. of Information Technology and Electrical Engineering, ETH Zürich, Zürich, Switzerland} \\
kuehnej@ethz.ch}}

\maketitle

\begin{abstract}

\emph{Optical Flow} (OF) is the movement pattern of pixels or edges that is caused in a visual scene by the relative motion between an agent and a scene. OF is used in a wide range of computer vision algorithms and robotics applications. While the calculation of OF is a resource-demanding task in terms of computational load and memory footprint, it needs to be executed at low latency, especially in robotics applications. Therefore, OF estimation is today performed on powerful CPUs or GPUs to satisfy the stringent requirements in terms of execution speed for control and actuation. On-sensor hardware acceleration is a promising approach to enable low latency OF calculations and fast execution even on resource-constrained devices such as nano drones and AR/VR glasses and headsets. This paper analyzes the achievable accuracy, frame rate, and power consumption when using a novel optical flow sensor consisting of a global shutter camera with an \emph{Application Specific Integrated Circuit} (ASIC) for optical flow computation. The paper characterizes the optical flow sensor in high frame-rate, low-latency settings,  with a frame rate of up to 88 fps at the full resolution of 1124 by 1364 pixels and up to 240 fps at a reduced camera resolution of 280 by 336, for both classical camera images and optical flow data.

\end{abstract}


\section{Introduction}

One important operation in visual perception pipelines is the tracking of image features using a descriptor-based approach. While descriptors allow the tracking of a single feature over arbitrary frames with no strict succession in time \cite{jiang2020efficient}, there is a subgroup of these algorithms that tracks features only over consecutive frames. This approach is called optical flow. It tracks the movement of single features in an image sequence or stream. Optical flow can also be applied in a dense manner to track the movement of every pixel in a frame \cite{blachut2022real}.

Tracking the movement of features between consecutive frames compared to feature matching in arbitrary scenes has the benefit that if the movement characteristic of the agent carrying the camera is known, then a maximum displacement can be defined to reduce the search range in feature matching \cite{delmerico2019we}. On the other hand, this task is confronted with some challenges compared to arbitrary feature matching. For example, if a feature is lost from one frame to the next, but appears in a later frame again, the inference that this is the same feature as was tracked before, would not be possible with only optical flow.

On resource-constrained hardware, we face additional challenges in the computation of OF due to the small form factor, a constrained energy budget, weight limitations, limited computing power, and often a requirement for low-cost solutions \cite{Palossi2019}. For instance, small-scale UAVs and AR/VR glasses and headsets, impose high requirements on their respective perception pipeline. On nano UAVs, a low-latency perception pipeline is equally important for fast flight as on its bigger counterparts \cite{Lu2018,merzlyakov2021comparison}. Similarly, compact and inconspicuous AR glasses also require a low-latency perception of the environment and wearer to appropriately display rendered information. 

This work introduces and characterizes a camera with an embedded OF accelerator that has the potential to address these problems, by offering low-latency OF tracking with moderate power requirements due to its domain-specific accelerator in the form of an ASIC.

The calculation of optical flow is a repetitive task that needs to be done several times per frame (usually for a fixed number of features) in a perception pipeline. This paper exploits a not yet commercially available optical sensor, designed by STMicroelectronics\footnote{www.st.com}, that embeds hardware acceleration for the computation of optical flow directly in the sensor. We present an accurate characterization of this sensor named VD56G3, evaluating its benefits in the low latency calculation of optical flow, which can reach up to 300 frames per second. The paper proposes and evaluates different use cases where this sensor might be useful, especially in the domain of high-speed autonomous vehicles and drones. Although we motivate the use of optical flow for small-scale UAVs we do not limit our analysis to UAVs. We also characterize the sensor in both high frame rate and large field of view settings that are more typical to other use cases such as speech recognition and eye-tracking. 

In detail, the paper contributes the following results:
\begin{itemize}
    \item A characterization of the VD56G3 sensor with different sensor settings on different application scenarios.
    \item An analysis of the tracking accuracy of the VD56G3 sensor in comparison to a Vicon ground truth.
    \item An analysis of the re-detectability of features across multiple frames.
    \item An investigation of the consumed power when running the optical flow pipeline on the camera, to assess if it is viable compared to running it on a microprocessor.
    \item A dataset containing both the optical flow and image data for different camera movements, as generated by the VD56G3 sensor.
\end{itemize}

\section{Related Work}
In this section, we discuss related work in the area of feature-based optical flow predictors by looking at feature detection, description, and matching algorithms. Furthermore, we look at the efforts that have been made to build hardware accelerators for the computation of optical flow.

\subsection{Feature Detection, Description, and Matching}

The detection of features is often achieved with a layered approach, where first a fast algorithm is run to get some candidate features. In the second step, a more sophisticated algorithm is used to determine if the candidate point is suitable as a feature point. These algorithms usually calculate the image gradient using either the Shi-Tomasi \cite{shi1994good} or the Harris \cite{harris1988combined} corner detector algorithm. If the image gradient is steep enough, then a point is selected as a feature and the feature descriptor is calculated. As a last step, other nearby feature candidates are being suppressed to avoid describing the same feature multiple times. Modern algorithms \cite{rublee2011orbfeature} use feature detectors that are rotation invariant as well as scaling invariant within a certain range by determining a dominant orientation and scale.

The feature descriptors, which are calculated on the de-rotated and re-scaled feature points utilizing the dominant orientation and scale respectively, are usually either composed of a binary gradient representation as in SIFT \cite{lowe2004sift}, BRIEF \cite{calonder2010brief}, or ORB \cite{rublee2011orbfeature}, or composed of the actual de-rotated and re-scaled image patch. 

The matching of features is done by finding the most similar correspondence between two feature descriptors of two different frames, which can be consecutive (as in the calculation of optical flow) or not (as in place recognition or key-frame-based VIO). As a similarity metric for binary descriptors, the Hamming distance is used. For image patches, the similarity between the template and the target patch is calculated using the normalized cross-correlation (NCC) or a similar metric \cite{Fraundorfer2012VO}.

\subsection{Optical Flow Prediction}
Optical flow describes the displacement of a pixel or a feature from one image frame to the next one. It can be therefore calculated by feature detection and matching as described previously. As an alternative method, it can be directly calculated by minimizing the photometric error, under the assumption that the brightness of the pixels in an image is nearly constant and the displacement between two images is small \cite{wang2018moving}. 

Additionally, the emergence of event cameras and the differential nature of event data has also led to research in optical flow prediction from event data. Both optical flow predictors on pure event data \cite{khairallah2022pca,gallego2018unifying} and predictors that use a combination of conventional camera frames and event data have been demonstrated \cite{lele2022fusing}. As conventional cameras and event cameras have some opposing strengths and weaknesses, for instance, optical flow from event frames suffers from drift, due to the integration of differential data, whereas conventional camera frames are unaffected by this drift, but exhibit a lower dynamic range and a lower frame rate than event camera, \cite{vidal2018ultimate} proposes to use a combination of event cameras, conventional cameras, and IMU data for the calculation of optical flow and general feature tracking.

%Application \cite{yue2022uav}.

\subsection{Hardware Acceleration for Optical Flow Prediction}
Computer vision tasks have been a popular target for hardware acceleration, especially stereo-vision on drones has been approached by various FPGA implementations. Although the computing platforms have become much more powerful ever since, the interest in smaller and smaller drones has led to continuous research in this area \cite{lu2021resource,wan2021survey}.

In the area of optical flow prediction, multiple accelerators have been developed in recent years. PX4FLOW a near-sensor accelerator, which consists of a camera and an ARM Cortex M4F micro-controller for optical flow prediction has been published almost a decade ago \cite{Honegger2013}. Recent works show FPGA implementations for the prediction of optical flow applying both traditional computer vision algorithms \cite{blachut2022real} as well as machine learning algorithms \cite{yan2023efficient}. Furthermore, \cite{stumpp2022harms} presents a hardware accelerator for the prediction of optical flow from event camera data. 

\begin{figure}[th]
    \centering
    \includegraphics[width=0.93\linewidth]{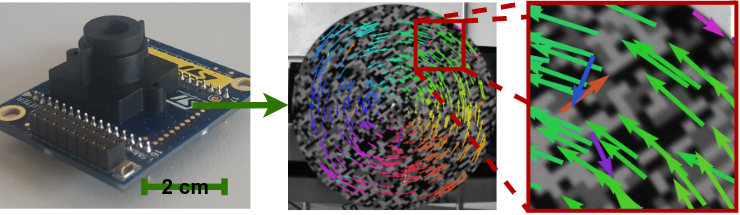}
    \caption{The VD56G3 image sensor includes an ASIC for the calculation of optical flow. The shown PCB on the left is part of the evaluation kit. The sensor returns the captured image and optical flow data as shown on the right.}
    \label{fig:image_sensor_evk}
\end{figure}

\section{Hardware Setup and Background}
In our characterization, we used the VD56G3 image sensor of STMicroelectronics depicted in Figure \ref{fig:image_sensor_evk}, which apart from the image sensor and optics also hosts an ASIC for the calculation of optical flow.

\subsection{The VD56G3 Image Sensor}
The VD56G3 is a 1.5-megapixel global shutter image sensor with integrated optical flow motion vector computation. The image sensor has an 1124-pixel by 1364-pixel resolution and produces monochrome images at up to 88 frames per second at the full resolution. At lower resolutions, the image sensor can reach frame rates as high as 300 frames per second. To get to lower-resolution images, the original image can be either cropped or sub-sampled, or a combination of both. The image sensor supports 2x and 4x sub-sampling and binning, as well as almost arbitrary cropping, through an area of interest setting. The optical flow unit can operate on input images up to VGA size (640 by 480 pixels), therefore the input image is automatically down-sampled if it is bigger than the supported size. 

The camera implements the MIPI CSI-2 interface, where either one or two data lanes can be used. Both the image data and the optical flow data are transmitted via this interface. The user can selectively only send image data, optical flow data, or both. Multiple regions of interest (ROI) can be defined, and with those settings, distinct image regions can be selected for the creation of the image that will be transmitted, the optical flow calculation, and the auto exposure controller. In our experiments, we kept all three ROIs identical.

\begin{table}[t]
    \centering
    \caption{Achievable frame rates for a given frame height and a given number of optical flow vectors when using the 10-bit ADC mode.}
    \begin{tabular}{c|c|c|c}
        Format & Frame Height [pixel] & \# OF Vectors & Frame Rate [1/s] \\
         \hline
        QVGA & 240 & 1024 & 338 \\
        QVGA & 240 & 2048 & 288 \\
        VGA & 480 & 0 & 229 \\
        VGA & 480 & 1024 & 205 \\
        VGA & 480 & 2048 & 186 \\
        FULL & 1364 & 0 & 88 \\
        FULL & 1364 & 1024 & 84 \\
        FULL & 1364 & 2048 & 80
    \end{tabular}
    \label{tab:framerates}
\end{table}

\subsection{Optical Flow Pipeline}
The VD56G3 camera employs the FAST algorithm as a feature detector and the BRIEF algorithm \cite{calonder2010brief} to generate oriented descriptors on one image level that either corresponds to the full image or a 2x subsampled image if the recorded resolution is bigger than VGA. Furthermore, the VD56G3 camera enforces a spread of descriptors across the entire ROI by limiting the number of descriptors per 16- by 16-pixel patch, this limit can be in the range of 2 to 8. 

The optical flow unit can predict up to 2048 motion vectors across the full image, which can be achieved by setting the desired number of BRIEF descriptors up to 2048. This setting and the actual number of produced BRIEF descriptors are then fed into a controller that regulates the contrast threshold of the FAST detector to reach the desired number of descriptors. After the matching with the BRIEF descriptors of the previous frame, while adhering to a maximum displacement setting, the number of effectively found matches, and therefore optical flow vectors, is typically significantly lower than the number of the BRIEF descriptors as not all the features can be matched. This can be due to certain features moving out of the visible frame, or becoming occluded, additionally, the feature appearance could also change significantly due to lighting changes, such that it can no longer be matched.

Using the Hamming distance, the best and second best match are determined and the displacement to the best fit as well as both Hamming scores are stored. The ratio between the two Hamming scores can be used by a later processing step, to suppress features that are not distinctive enough \cite{lowe2004sift}. One optical flow vector is represented with the following six values: the x and y coordinates of the feature in the previous frame, the difference in coordinates from the previous to the current frame, and the best and second best Hamming score. This information is then transmitted via MIPI CSI-2 in lines of 16 optical flow vectors.

\label{sec:cam_limits}
In Table \ref{tab:framerates} we indicate the achievable frame rate for a given frame height and a given number of optical flow vectors in the 10-bit ADC mode, which ranges between 80 and over 300 frames per second.

\section{Dataset}
To be able to evaluate the VD56G3 sensor, we created a dataset that contains linear and rotational movement patterns. To better understand the effect of separate movement directions, we built a setup that creates movement in one of the degrees of freedom of the camera. Furthermore, to assess the tracking accuracy in different scenarios, we used multiple images and color patterns to generate features for the optical flow unit of the camera.

\begin{figure}[t]
    \centering
    \includegraphics[width=0.89\linewidth]{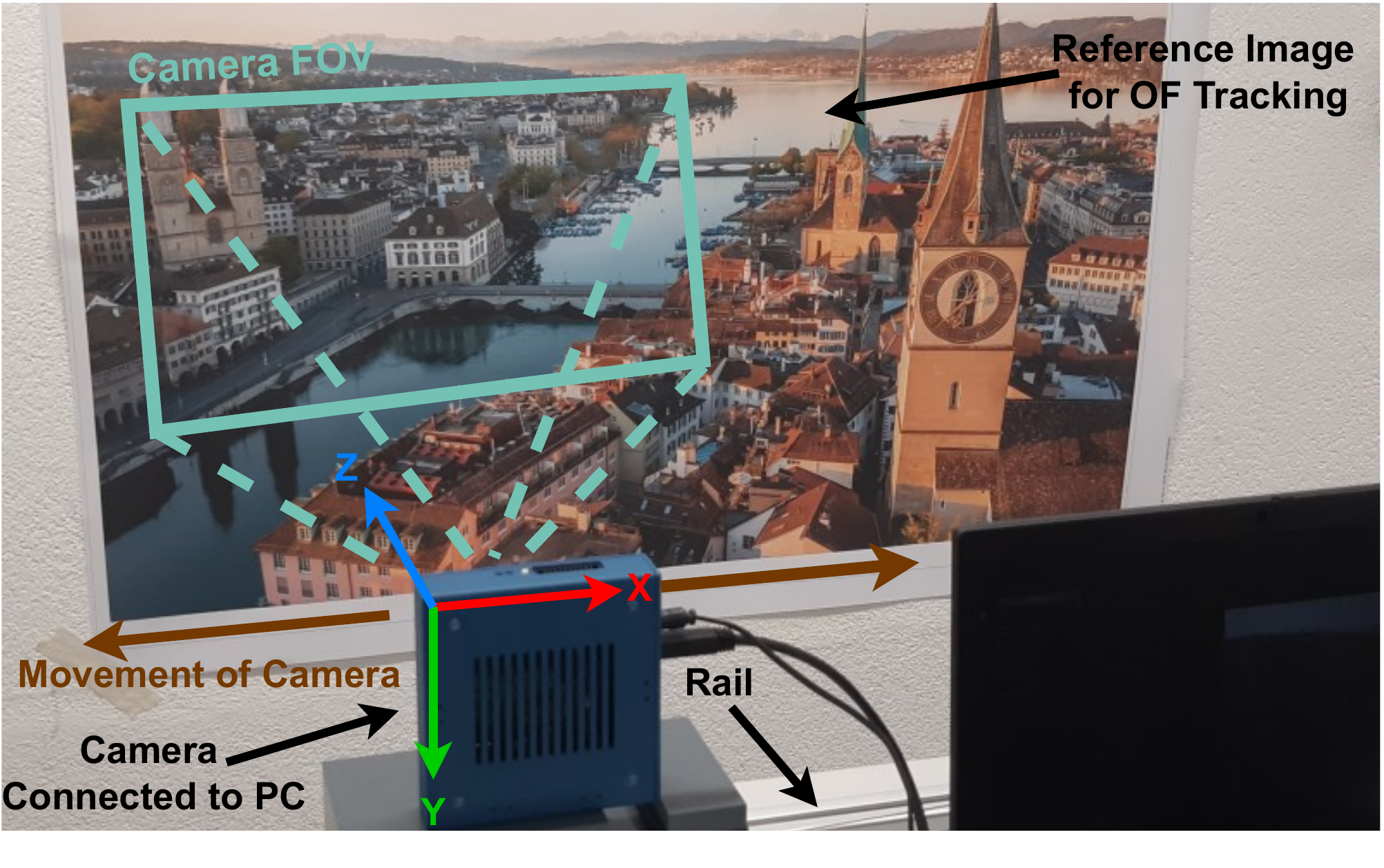}
    \caption{The setup that was used for the recording of the sequences where the camera moves parallel to the image plane. To make sure the optical flow camera has enough features to track, two different printed images were used.}
    \label{fig:dataset_setup}
\end{figure}

\begin{table*}[t]
    \centering
    \caption{Parameter sets of camera settings that were applied during the recording of the dataset.}
    \begin{tabular}{c|c|c|c|c|c|c|c}
        Parameter Set & Camera Resolution X & Camera Resolution Y & Crop$^{\mathrm{a}}$/Sub-sample & Frame Rate & BRIEF Target & BRIEF Max & OF Spatial Point  \\
        \hline
        1 & 1124 pixel & 1364 pixel & None & 60 FPS & 1536 & 2048 & 2 \\
        2 & 1120 pixel & 1344 pixel & Crop (0,0) & 60 FPS & 1536 & 2048 & 2 \\
        3$^{\mathrm{b}}$ & 640 pixel & 480 pixel & Crop (240,432) & 140 FPS & 768 & 1024 & 4 \\
        4 & 560 pixel & 672 pixel & Crop (280,336) & 140 FPS & 768 & 1024 & 4 \\
        5 & 560 pixel & 672 pixel & 2x sub-sample & 140 FPS & 768 & 1024 & 4 \\
        6 & 272 pixel & 336 pixel & Crop (420,504) & 240 FPS & 384 & 512 & 8 \\
        7 & 280 pixel & 336 pixel & 4x sub-sample & 240 FPS & 384 & 512 & 8 \\
        \multicolumn{8}{l}{}\\
        \multicolumn{8}{l}{$^{\mathrm{a}}$If cropping is applied the numbers in brackets indicate the top left corner of the cropped image relative to the full 1124 by 1364 image.} \\
        \multicolumn{8}{l}{$^{\mathrm{b}}$Operate the optical flow unit at its maximum resolution.}
    \end{tabular}
    \label{tab:paramter_sets}
\end{table*}

\subsection{Test Case}
For the recording of the dataset, we defined four different movement patterns:
\begin{itemize}
    \item linear movement parallel to a planar scene,
    \item linear movement away from a planar scene,
    \item rotation around the focal axis,
    \item stand-still.
\end{itemize}

To ensure that the camera has enough features to track and to evaluate the repeatability of the feature detector we used printed images in A0-size. For the translation experiments, we used an image of a city as an example of easily distinguishable features, and an image of a forest with foliage as an example of features that are hard to track. For the rotation experiments, we used two different color-patterned wheels. For the linear movements, the camera was moved, for the rotational experiments we rotated the pattern at variable speeds instead. The setup where the camera moves parallel to the image plane is depicted in Fig. \ref{fig:dataset_setup}. For every combination of camera movement, image for the feature generation, and camera setting we recorded a 10-second sequence of optical flow and image data.

\subsection{Image Sensor Settings}
We show a wide range of settings that cover different potential use cases such as visual (inertial) odometry that requires a large field of view in combination with a moderate frame rate, but also eye tracking for example for the use in virtual or augmented reality glasses, which requires a small field of view in combination with a fast frame rate to track the eye movements. The parameter sets are given in Table \ref{tab:paramter_sets}.

As the image sensor has a multitude of possible settings that can be combined in various ways, seven fixed settings were chosen for our experiments. Those seven sets of settings were all selected close to the maximum resolution for a given frame rate as indicated in Table \ref{tab:framerates} while keeping a fixed aspect ratio. The dimensions indicated in Table \ref{tab:paramter_sets} describe the resulting resolutions of the output image after sub-sampling or cropping. We select a target for the number of BRIEF descriptors (\emph{BRIEF Target}) that is roughly scaled proportionally to the image length. Furthermore, if more BRIEF descriptors than specified in the \emph{BRIEF Max} setting would be generated, those descriptors are dropped. As the feature detector processes an image frame from top to bottom, this can result in a lack of features in the bottom part of the image frame. The \emph{OF Spatial Point} setting indicates how many features shall be detected within every 16 pixels by 16 pixels patch. With this setting, one can either enforce a more spread-out distribution of feature points or a high concentration on very distinctive features depending on the application.

It is important to note that in parameter sets 1 and 2 the optical flow output is calculated on a 2x down-sampled image as the original image has a width bigger than 640 pixels. As a reference, we added parameter set 3 which is exactly 640 pixels wide and therefore utilizes the maximum width that can be processed by the optical flow unit without down-sampling the image.

\section{Experimental Setup and Characterization}
In the following, we are describing different test scenarios that were used to verify the accuracy of the hardware-accelerated optical flow prediction compared to offline optical flow calculation.

\subsection{Dataset Validation}
\label{sec:dataset_validation_with_vicon} 
During the acquisition of the linear movement dataset parallel to the image plane, we additionally recorded the absolute position of the camera setup using a Vicon motion capture system\footnote{https://www.vicon.com/}. To evaluate the tracking accuracy of the VD56G3 sensor, we compare the actual displacement of the features in the scene to the estimate of the optical flow unit.

Additionally, we use the standstill dataset to have a look at the length of the feature tracks, i.e. the concatenation of the optical over multiple frames flow for a specific feature. We check for how many frames a feature can be tracked and if it is re-detected in a later frame.

\subsection{Power Profiling}
As a last experiment, we measure the power utilization for three extreme cases of the parameter sets defined in Table \ref{tab:paramter_sets} namely parameter sets 1, 4, and 6. For better comparison, we additionally run the parameter sets 4 and 6 at 60 frames per second. For all cases, we analyze the power draw when optical flow is being calculated and compare it to the case where the optical flow calculation is turned off.

The camera sensor is supplied by three voltage rails, an analog supply with 2.8 volts, a digital supply with 1.8 volts, and a supply voltage for the sensor core with 1.2 volts. The specific hardware variant of the VD56G3 sensor that we used (the module of the evaluation kit) hosts an onboard LDO for the 1.2-volt supply. We therefore only measured the current on the 2.8-volt and 1.8-volt supplies.

\section{Experimental Results}
This section presents the accurate characterization of the compact and low-power optical flow camera, in terms of tracking accuracy, tracking repeatability, and power draw.

\begin{figure*}[ht]
    \centering
    \includegraphics[width=0.95\linewidth]{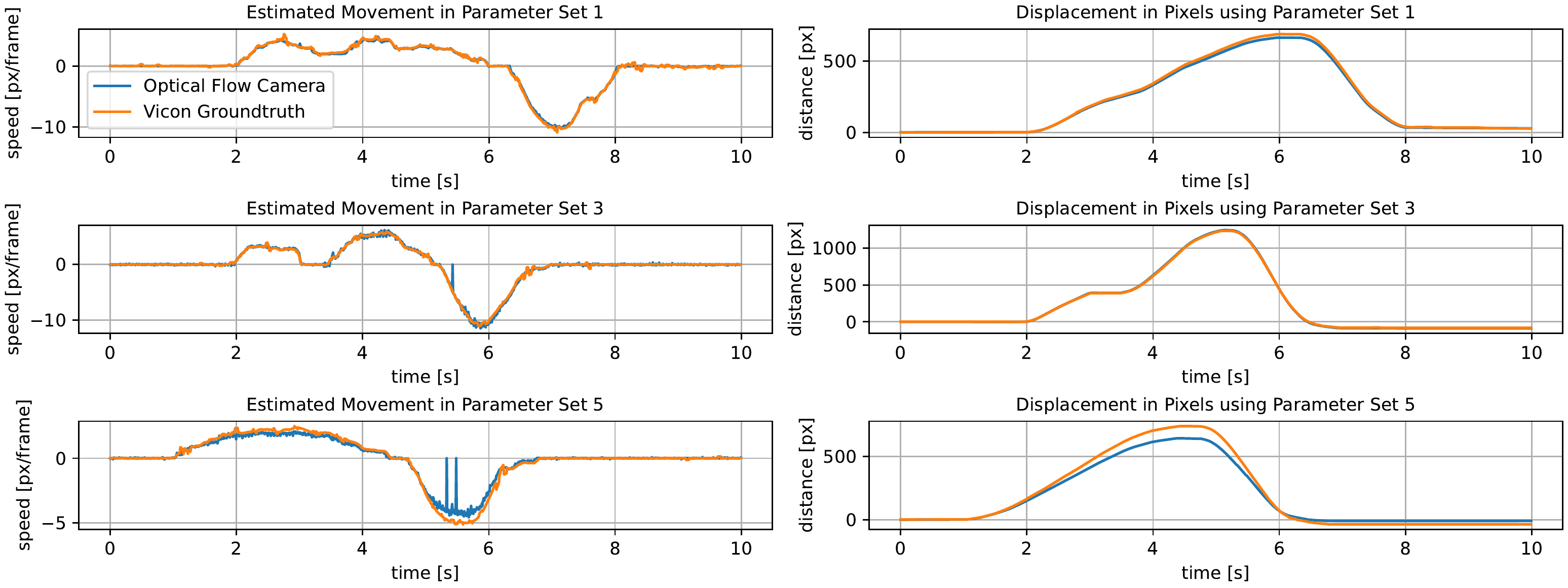}
    \caption{The plots on the left show a comparison of the tracked movement speed when moving the camera parallel to the image plane for the parameter sets 1, 3, and 5 when utilizing the image of the forest to generate features. The plots on the right show the total traveled distance in pixels of both the estimate from the optical flow camera and the Vicon ground truth.}
    \label{fig:linear_parallel_forest_selected}
\end{figure*}

\subsection{Accuracy}
The tracking accuracy was quantitatively analyzed by comparing the Vicon ground truth to the average optical flow that was produced by the VD56G3 sensor. The results of the linear movement parallel to the scene for the forest image are given in Figure \ref{fig:linear_parallel_forest_selected}. Generally, one can observe, that the tracking is most accurate when the camera is configured to have a large resolution and the frame rates and sub-sampling factors are selected, such that the average optical flow is above one pixel. As the resolution in parameter set 1 is bigger than VGA resolution, it is 2x subsampled before the optical flow is predicted, whereas the image in parameter set 3 is not subsampled, therefore although the frame rate of parameter set 1 is 60 frames per second and 140 frames per second for parameter set 3 the average movement speeds of the optical flow vectors are almost the same. From the plot in Figure \ref{fig:linear_parallel_forest_selected}, we can see that the parameter set 1 has slightly lower tracking accuracy which is caused by the loss in precision by the subsampling process. In parameter set 5, where both subsampling and a high frame rate of 140 frames per second are applied, one can observe an even bigger decrease in tracking accuracy. The remaining paramter sets confirm the previous findings, with parameter set 7 performing worst in the used scenario due to both the high frame rate and subsampling factor of 4x.

%\begin{figure}[t]
%    \centering
%    \includegraphics[width=\linewidth]{images/feature_track_hist.pdf}
%    \caption{Distribution of feature track lengths generated by the optical flow camera and OpenCV's ORB implementation respectively.}
%    \label{fig:feature_track_hist}
%\end{figure}

\subsection{Feature Track Length}
In a standstill image with constant illumination, the repeatability of the feature tracker of the optical flow unit is slightly lower than in a reference implementation where the ORB descriptor is used. Both descriptors are able to produce long feature tracks with lengths of 30 frames and above, whereas the density of long feature tracks generated by the ORB tracker is higher than the one of feature tracks generated by the tracker of the optical flow sensor. It is worth noting, that the reference ORB implementation iteratively optimizes the threshold to select corners for a given frame, whereas the equivalent threshold of the optical flow sensor is only changed between two frames.

\subsection{Power Profile}

\begin{table*}[th]
    \centering
    \caption{Current draw of the 1.8 V and 2.8 V power rail respectively as well as the total power draw for the selected parameter sets.}
    \begin{tabular}{r|c||c|c||c|c||c|c}
        \multicolumn{2}{c||}{} & \multicolumn{2}{c||}{1.8 V Power Rail} & \multicolumn{2}{c||}{2.8 V Power Rail} & \multicolumn{2}{c}{Total Power Draw} \\
        FPS & Resolution & Optical Flow Off & Optical Flow On & Optical Flow Off & Optical Flow On & Optical Flow Off & Optical Flow On \\
        \hline
         60 & high   & 64.10 mA & 88.00 mA & 21.94 mA & 21.95 mA & 176.81 mW & 219.86 mW  \\
         60 & medium & 56.90 mA & 76.23 mA & 18.17 mA & 18.19 mA & 153.30 mW & 188.15 mW  \\
         60 & low    & 53.77 mA & 68.80 mA & 16.32 mA & 16.37 mA & 142.48 mW & 169.68 mW  \\
        140 & medium & 65.73 mA & 93.79 mA & 23.84 mA & 23.91 mA & 185.07 mW & 235.77 mW  \\
        240 & low    & 58.20 mA & 77.74 mA & 23.65 mA & 23.76 mA & 170.98 mW & 206.46 mW
    \end{tabular}
    \label{tab:power_profile}
\end{table*}

We measured the 1.8 V and the 2.8 V supplies for the image sensor. We conducted the analyses on the parameter sets 1, 4, and 6 of Table \ref{tab:paramter_sets} and additionally ran the parameter sets 4 and 6 at 60 frames per second, both with the optical flow prediction enabled and disabled. We ran the current measurements again for 10 seconds. When streaming data we arrive at the average current draw for the 1.8 V and 2.8 V supply as given in Table \ref{tab:power_profile}, additionally, the resulting power consumption is also shown.

The 1.8 V supply is used for the digital logic in the camera, therefore, we can see a clear increase in current draw, when the optical flow unit is turned on. Furthermore, the current draw on the 1.8 V supply depends on the frame rate and the total number of pixels being processed per second. For the 60 FPS high-resolution case, the camera needs to process 92 Megapixels of image data per second, whereas, in the 240 FPS low-resolution case, only 22 Megapixels of image data are being processed per second.

On the 2.8 V supply, which is used for supplying the analog components, like the pixel array, we can see no significant change in the current draw when enabling the optical flow unit. When operating the camera at higher frame rates, we can see a very light increase in the current draw.

For the overall power draw, we can see an increase of 24.3\% (43.05\,mW) when enabling the optical flow unit in the 60 FPS high-resolution case. A slightly lower 20.8\% (35.48\,mW) increase in power draw is measured for the 240 FPS low-resolution case. We can observe, that the change in frame rate has almost no impact on the overall power consumption. One has to note that with both settings we are operating the camera close to the maximum specification.

\section{Discussion}
The qualitative rotation experiments show the drawback of a non-rotation-invariant feature descriptor. Rotations around the focal axis can only be tracked if they are sufficiently small, or the camera frame rate is sufficiently high. Otherwise, the descriptors of the rotated features are altered too much, such that they can no longer be matched.

The experiments show, that the frame rate of the sensor needs to be matched with the movement speed of the scene to obtain satisfying results. If the scene is moving, but the movement is below a pixel per frame, poor tracking accuracy is reached. For optimal results, one can adjust the frame rate and the sub-sampling factor, depending on the average movement speed of the scene. As the sensitivity of the feature detector is updated only once per frame and not iteratively on a single frame, it sometimes fails to detect features at all, which results in a movement speed prediction of zero in Figure \ref{fig:linear_parallel_forest_selected} for parameter sets 3 and 5.

If we compare the added power draw of the optical flow unit of 43.05\,mW against other low-power, high-frame-rate optical flow implementations, such as \cite{Honegger2013}, which is an MCU implementation with a similar power envelope targeted at drone applications, we can confirm that the VD56G3 sensor is suitable for drone applications. Furthermore, thanks to its ASIC implementation of a feature-based optical flow estimator the VD56G3 sensor produces significantly more optical flow data than the implementation proposed in \cite{Honegger2013}, which applies patch matching on a 64- by 64-pixel image at up to 250 frames per second.

\section{Conclusion}
This paper presented and characterized a novel optical flow camera that leverages on-sensor acceleration. With accurate experimental evaluation, the paper demonstrated the low latency, the low power, and the accuracy of the optical flow prediction. Furthermore, guidelines on optimal configuration and indications on how to adapt them are provided. We have shown the potential in low-latency applications and especially for low-power edge applications, where strict energy constraints and fast movements are demanding an energy-efficient and low-latency solution.

\bibliographystyle{IEEEtran}
\bibliography{references}

\end{document}